\RequirePackage{fix-cm}   
\documentclass[12pt,a4paper]{iopart}
\pdfoutput=1

\usepackage{tikzexternal}   
\tikzset{external/only named={true}}   
\newcommand{\Ecuts}[1]{\pgfmathsetmacro{\Ecuthi}{#1*46.366} 
\pgfmathsetmacro{\Ecutlo}{\Ecuthi+147}}  
\def\singleplotwidth{0.5\linewidth}
\def\stepplotwidth{0.38\linewidth}\def\stepplotheight{0.32\linewidth}

\usepackage{microtype}
\usepackage{fixltx2e}

\expandafter\let\csname equation*\endcsname=\relax
\expandafter\let\csname endequation*\endcsname=\relax
\usepackage[fleqn]{amsmath}   
\newcommand{\eval}[2][\right]{\relax
  \ifx#1\right\relax \left.\fi#2#1\rvert}
\newcommand{\dd}{\mathrm{d}}

\usepackage{url}   
\usepackage[breaklinks,pdfborder={0 0 0}]{hyperref}

\begin{document}

\title[Diffusion-limited reactions on disordered surfaces]
{Diffusion-limited reactions on disordered surfaces with
  continuous distributions of binding energies}
\author{A Wolff$^1$, I Lohmar$^2$, J Krug$^1$ and O Biham$^2$}
\address{$^1$ Institute for Theoretical Physics, University of Cologne,
  Z\"ulpicher Str. 77, 50937 K\"oln, Germany}
\ead{awolff@thp.uni-koeln.de}
\address{$^2$ Racah Institute of Physics, The Hebrew University, Jerusalem
  91904, Israel}

\begin{abstract}
  We study the steady state of a stochastic particle system on a
  two-dimensional lattice, with particle influx, diffusion and
  desorption, and the formation of a dimer when particles meet.  Surface
  processes are thermally activated, with (quenched) binding energies
  drawn from a \emph{continuous} distribution.  We show that sites in
  this model provide either coverage or mobility, depending on their
  energy.  We use this to analytically map the system to an effective
  \emph{binary} model in a temperature-dependent way.  The behavior of
  the effective model is well-understood and accurately describes key
  quantities of the system: Compared with discrete distributions, the
  temperature window of efficient reaction is broadened, and the
  efficiency decays more slowly at its ends.  The mapping also explains
  in what parameter regimes the system exhibits realization dependence.

\end{abstract}

\submitto{Journal of Statistical Mechanics: Theory and Experiment}

\section{Introduction}\label{sec:intro}

The interplay of diffusive transport and quenched random reaction rates
poses some of the most intriguing problems in the statistical physics of
disordered systems~\cite{benavraham00}.  While the simple case of
single-particle diffusion in random media is reasonably well
understood~\cite{haus87-diffus-regul-disor-lattic,
  bouchaud90-anomal-diffus-disor-media}, already the linear dynamics
that arises from adding an autocatalytic reaction term to the diffusion
(heat) equation generates complex, intermittent spatio-temporal
patterns~\cite{zeldovich87-inter-random-media} that have only recently
become tractable by rigorous
analysis~\cite{gaertner06-inter-catal-random-medium}.

In the present paper we consider a particular variant of this general
class of problems, which is motivated by the physics of heterogeneous
catalysis on disordered
surfaces~\cite{zhdanov02-monte-carlo-simul-oscil-chaos, frachebourg95,
  head96-kinet-catal-surfac-disor,
  oshanin04-exact-solvab-model-monom-monom}.  We study a large (but
finite) two-dimensional lattice system with stochastic particle
dynamics, including influx, desorption, diffusion and pairwise reaction
upon meeting.  The rates of desorption and diffusion are subject to
quenched disorder.  An important realization of this type of dynamics
are chemical reactions on dust grains in interstellar clouds, for a
review see~\cite{herbst05-chemis-inter-grain}.  In the paradigmatic
reaction in this context, hydrogen atoms from the gas phase collide with
and stick to a dust grain, they diffuse on its surface, and if two of
them meet, they form an $\mathrm{H}_2$ molecule~\cite{gould63,
  hollenbach71a}.  The key quantity of such systems is their
steady-state \emph{efficiency}, i.e., the fraction of incoming particles
which leave due to the reaction (as opposed to thermal desorption of a
particle before it takes part in a reaction)---the significance for
applications is evident.  All other parameters being fixed, this is
typically a function of the system temperature, and high efficiency is
limited to a specific temperature range.  Below this range, particles
become immobile and can no longer react, while above they are thermally
emitted too quickly.

In previous work, we and others have first studied the system with
homogeneous rates for all processes, when one can obtain analytical
results~\cite{green01, biham02, lohmar06, lohmar08,
  lohmar09-accur-rate-coeff}.  However, spatial inhomogeneities in the
process rates are of theoretical interest and of importance for
applications.  We have therefore started a systematic analysis of the
effect of disorder in the local rates of hopping and desorption.
In~\cite{wolff10-react-two-dimen-bin-diso}, we considered a
\emph{binary} model that consists of a lattice of adsorption sites, each
associated with one of two possible binding energies, and labeled as
standard (``shallow'') and strong-binding (``deep'') sites,
respectively.  All effects on the efficiency seen in kinetic Monte Carlo
(KMC) simulations of this system have been well-understood, and we have
provided thorough explanations in terms of microscopic processes,
complemented by analytical mean-field results.  This \emph{binary} case
is important both as a starting point for theory as well as for
applications, where one can often naturally identify shallow and deep
sites.
Note that even in this case, there are no exact analytical results
beyond a mean-field description (particularly not for finite system
size).

The generalization to the more generic case of \emph{continuous}
distributions of binding energies now naturally suggests itself for
applications, and it is of fundamental interest to the theory of
disordered systems, where it is well known that the nature of the
disorder distribution (i.e., discrete vs.\ continuous) may
\emph{qualitatively} affect the behavior.  For example, for the random
field Ising model~\cite{imry75-random-field-instab,
  nattermann98-theor-random-field-ising-model}, a bimodal distribution
of the local field strength has been suggested to give rise to a
first-order phase transition, whereas other, continuous, distributions
do not~\cite{swift97-scalin-random-field-ising-model} (see
also~\cite{fytas08-first-order-trans-featur-3d} and references therein
for an account of the still ongoing debate).  Similarly, in the context
of random Schr\"odinger operators the shape of the Lifshitz tails in the
electronic density of states is known to differ markedly between
discrete and continuous (bounded) disorder
distributions~\cite{luck88-lifsh-tails-long-time-decay}.

In this paper we extend the analysis of diffusion-limited surface
reactions from the binary case~\cite{wolff10-react-two-dimen-bin-diso}
to continuous distributions.  The central result of our work is the
existence of a simple mapping from the case of binding energies drawn
from a continuous distribution to an \emph{effective} binary model as
described above.
This mapping is highly intuitive and we argue for its validity supported
by strong numerical evidence.
The binary model in turn is well-understood and, using mean-field
methods, easily soluble (analytically in principle, and using minimal
numerical techniques in practice) for the steady-state coverage and
efficiency.
Note that the mapping depends on the system temperature, which is
responsible for the fact that the system's behavior is qualitatively
different from the genuinely discrete distribution case studied
previously.
Together we thus provide both a detailed understanding of the physics of
stochastic particle systems with disordered binding energies as well as
a straight-forward way to calculate key quantities of practical
interest.
Moreover, it is a question of general importance for the analysis of
stochastic particle systems if (and how) a continuous distribution may
be effectively replaced by a simpler discrete one.  For the particular
system studied here, we answer this question comprehensively (and
affirmatively).  The accessible microscopic explanations may also
provide hints for the possibility of such a mapping for related models
from other fields.

The paper is organized as follows.  In \sref{sec:model-and-sim} we
introduce the model and define the notation and terminology.  We also
describe the simulation techniques and the parameters used.  We give a
brief review of the necessary background for the corresponding
homogeneous and binary systems in \sref{sec:review}.  In
\sref{sec:effective} we present the effective binary model and derive
the mapping to it.  We show the excellent agreement of the model results
with those of KMC simulations.  We also analyze the shape of the
efficiency tails and the issue of sample-to-sample fluctuations.
Lastly, we summarize our findings and their implications in the
conclusions (\sref{sec:conclusions}).

\section{Model and simulation}
\label{sec:model-and-sim}

\subsection{Definition of the model}
\label{sec:definition-model}

The model surface is a two-dimensional square lattice of $S$ sites with
periodic boundary conditions.
On this lattice we consider the dynamics of particles of a single
species.  They impinge onto the lattice at a homogeneous rate $f$
per site.  If a site is already occupied, the impinging particle is
rejected.  In the context of surface chemistry this is known as
Langmuir-Hinshelwood (LH) rejection~\cite{langmuir18}.

Particles explore the lattice by hopping to neighboring sites with an
(undirected) rate $a_s$, and they can spontaneously leave a site by
desorption with rate $W_s$.  Both rates depend on the current particle
position $s$ (but not on its neighborhood).  If two particles meet on
one site, they react to form a dimer and leave the system immediately.
We denote the total rate of these reaction events in the system by $R$.
The key quantity of such a system is the \emph{efficiency} $\eta$,
defined as the ratio between the number of particles that react and the
total number of impinging particles, when the system is in a steady
state,
\begin{equation}
  \eta = 2 R / (f S).
\end{equation}
We denote the steady-state number of particles on the grain by $N$, such
that the \emph{coverage} reads $\theta = N/S$.

In view of possible applications (cf.\ \sref{sec:intro}), we choose the
rates $W_s$ and $a_s$ to be thermally activated by a system temperature
$T$.  The activation energy for desorption or \emph{binding energy} at
site $s$ is denoted $E_{W_s}$.  Similarly, hopping from that site has an
activation energy $E_{a_s}$.
All rates share the attempt frequency $\nu$, so that
\begin{equation}
  W_s = \nu \exp \left(-\frac{E_{W_s}}{T}\right),
  \qquad\text{and}\quad
  a_s = \nu \exp \left(-\frac{E_{a_s}}{T}\right).
\end{equation}
Here and in the following energies are measured in temperature units.

\begin{figure}
  \centering
  \tikzsetnextfilename{fig1}%
  \begin{tikzpicture}[scale=0.65,
    to path={.. controls +(+0.25,0) and +(-0.25,0) .. (\tikztotarget)
      \tikztonodes}]
    \pgfmathsetmacro{\Ea}{2}
    \pgfmathsetmacro{\wid}{0.8}
    \pgfmathsetmacro{\Edes}{1.5}
    \pgfmathsetmacro{\last}{9}
    \draw[<->] (\last+1,\Edes) node[right] {$s$}
    -| (0,-\Ea-\wid-0.5) node[left] {$E_{W_s}$};
    \draw[<->] (1.5,0) -- +(0,\Edes) node[right,midway]
    {$E_{W_s}-E_{a_s} =\Delta E$};
    \draw[sketch level] (0,0) -- (\last+1,0);
    \draw[sketch level] (0,-\Ea) node[left] {$\langle E_W\rangle$} -- +(\last+1,0);
    \foreach \i in {0,...,\last} 
    \draw[main graph] (\i,0) to (\i+0.5,-\Ea-rand*\wid) to (\i+1,0);
    \path[use as bounding box] (-1,\Edes+0.5) rectangle (\last+2,-\Ea-\wid-0.5);
  \end{tikzpicture}
  \caption{One-dimensional cut through the energy landscape of our
    model.}\label{fig:E-scape}
\end{figure}
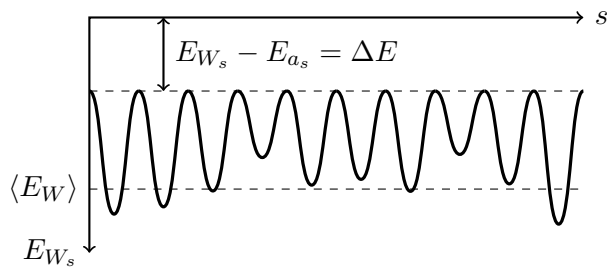

In principle, $E_{a_s}$ could be independent of $E_{W_s}$, but we want
to ensure \emph{detailed balance}.  The simplest way to achieve this is
by choosing $W_s/a_s = \mathrm{const}$, and we will employ this choice
throughout.  Equivalently, $E_{W_s}-E_{a_s} = \Delta E$, where $\Delta
E$ is a constant (independent of the site) describing the additional
energy needed for desorption on top of the local transition state
energy.  The average number of sites visited by a single particle before
desorption becomes then independent of the local rates and $\approx
a_s/W_s$ (for $W_s\ll a_s$).
Finally, the binding energy $E_{W_s}$ for each lattice site $s$ is drawn
once from a probability density function (PDF) $\rho(E_W)$ and remains
fixed (quenched disorder realization).  A one-dimensional cut through
such an energy landscape is sketched in \fref{fig:E-scape}.  Without
desorption, this construction of the energy landscape and the associated
hopping rates is known as the ``random trap
model''~\cite{haus87-diffus-regul-disor-lattic}.

\subsection{Kinetic Monte Carlo simulation}
\label{sec:kmc}

Our reference point for the system behavior is provided by extensive
kinetic Monte Carlo simulations.  The standard algorithm proceeds as
follows (cf.~\cite{voter07} for a review).  We keep track of the full
microscopic dynamics of continuous-time random walkers~\cite{montroll65}
with standard exponential waiting time distributions.  In each
simulation step, the current system configuration determines the list of
possible elementary processes (influx, as well as desorption and
nearest-neighbor hopping of all particles on the lattice) and their
rates.  By comparing a random number with the normalized partial sums of
these rates we find the process to execute next.  The simulation time is
then advanced according to the total sum of rates and the configuration
is updated.

For a given set of parameters, each realization of the model is
characterized by the set of binding energies for all sites, which are
independently drawn from a distribution.
As characteristic examples of different types of distributions we
consider a) the uniform distribution over a certain interval, with
bounded support, b) the (shifted) exponential distribution, which has a
low-energy cutoff, but a high-energy tail, and c) the normal
distribution, with tails to low as well as high energies.  The latter
case is often appropriate for the description of systems where binding
energies are affected by many different random influences.

Unless indicated otherwise, we simulate $10$ realizations for each set
of parameters; and for each set, these realizations are drawn anew.  The
different realizations are not used for averaging, but rather so we can
examine sample-to-sample variations due to quenched disorder.  For a
given realization, we wait for the system to reach the steady state.  We
then record the efficiency and the spatial distribution of reaction
events, as well as the local and global coverage, over $10^8$
impingements.

For easy comparison, the model parameters are mostly taken from the
astrophysical application mentioned before, as in our earlier
work~\cite{wolff10-react-two-dimen-bin-diso}.
This guarantees that we observe interesting kinetic regimes, and, as a
side effect, it highlights the relevance of our findings for
applications.
We use a quadratic square lattice of $S=100\times 100$ sites.  The
particle influx per site is $f = 7.3\times10^{-9}\ \mathrm s^{-1}$ (as
obtained from typical values for the density and the thermal velocity of
hydrogen atoms in interstellar clouds and from the density of adsorption
sites on an amorphous carbon sample~\cite{biham01}).
For the attempt frequency we choose the standard value of $\nu=10^{12}\ 
\mathrm s^{-1}$ which is commonly used throughout surface science.
The energy landscape of the surface is described by the mean binding
energy $\langle E_W\rangle = 658\ \mathrm K$, and the mean diffusion
barrier $\langle E_a\rangle = 511\ \mathrm K$ (as found for hydrogen
atoms on amorphous carbon~\cite{katz99}), with the difference $\Delta E
= 147\ \mathrm K$ being constant for all sites.  The standard deviation
of \emph{both} energy distributions is denoted $\sigma$; we often use
the value normalized by the mean binding energy, $\tilde \sigma = \sigma
/ \langle E_W\rangle$, which we vary between $10\%$ and $50\%$.

KMC simulations of similar and of more complex models have been
performed in the astrophysical context~\cite{chang05}.  Additional
features include, among others, stochastic heating of the system, and
accounting for the surface morphology (implying strongly correlated
binding energies of adjacent sites), see,
e.g.,~\cite{herbst06-monte-carlo-studies-surfac-chemis}.
In contrast, the present work is not so much concerned with the concrete
temperature ranges of efficient reaction for a certain set of
parameters, but we rather strive for a coherent explanation of the
underlying physics of the generic system.

\section{Review of homogeneous and binary systems}
\label{sec:review}

In the remainder of this article we will repeatedly use concepts and our
understanding of the homogeneous and binary case.  Therefore we start
with a brief review of these systems.  We will consider steady-state
conditions exclusively.

\subsection{Homogeneous system}
\label{sec:homogeneous}

\begin{figure}
  \centering
  \tikzsetnextfilename{fig2}%
  \begin{tikzpicture}
    \begin{axis}[
      xlabel=Temperature $T/\mathrm K$
      ,ylabel={\color{blue}Cvg.\ $\theta$, \color{red}Eff.\ $\eta$}
      ,width=\singleplotwidth
      ,xmin=8,xmax=21,ymin=0,ymax=1.05
      ,minor tick num=4
      ,x tick label style={/pgf/number format/precision=0}
      ,y tick label style={/pgf/number format/precision=1}
      ]
      \addplot[main graph,blue,solid] plot
      file {NdST-binre-750longhop-0.dat};
      \addplot[main graph,red,solid] plot
      file {etaT-binre-750longhop-0.dat};
    \end{axis}
  \end{tikzpicture}
  \caption{Coverage $\theta$ (blue) and efficiency $\eta$ (red) as
    functions of temperature $T$, in the homogeneous model described by
    a rate equation.  Parameters are chosen as described in
    \sref{sec:kmc}, using the mean values of activation energies.}
  \label{fig:cvgetahom}
\end{figure}
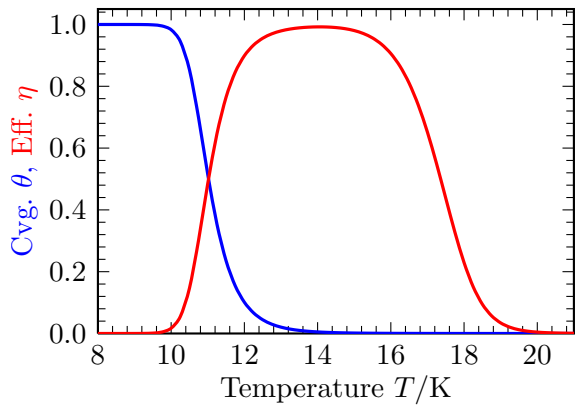

The homogeneous system is characterized by a single binding energy
$E_W$ and a single hopping activation energy $E_a$
for all sites.  It can be solved analytically using rate equations (and
including LH rejection)~\cite{katz99,biham01}, or with the master
equation (not including LH rejection) accounting for fluctuations as
well~\cite{green01, biham02}.  Results for coverage and efficiency,
obtained using the rate equation approach (which is accurate for
sufficiently large systems), are shown in \fref{fig:cvgetahom}.  The
efficiency is limited to a narrow window of temperatures, while the
coverage monotonically decreases with increasing temperature.  Note that
while the coverage is very small above $T = 14\ \mathrm{K}$, the
reaction remains efficient as long as the mean number of sites the
particles visit is still larger than the mean number of empty sites
surrounding each particle.

For later reference we recall the efficiency obtained in the rate
equation approach~\cite{biham02},
\begin{equation}\label{eta-hom}
  \eta = 1 - \frac{(W+f)^2}{4fa}
  \left(\sqrt{1+\frac{8fa}{(W+f)^2}} - 1\right).
\end{equation}
This implies temperature bounds for the window of efficient reaction,
where $\eta \geq 1/2$:
\begin{equation}
  \label{Tlow}
  T^\mathrm{low}=\frac{E_a}{\ln(\nu/f)} 
\end{equation}
is the temperature
below which particles arrive faster than they hop.  Hence, the coverage
is close to unity, leading to dominant LH rejection and low efficiency.
Similarly,
\begin{equation}
  \label{Tup}
  T^\mathrm{up}=\frac{2 E_W-E_a}{\ln(\nu/f)}
\end{equation}
is the temperature
above which desorption ends the typical particle residence before it can
react.  Therefore, in addition to the very low coverage, the
efficiency is low as well.
The temperature of maximal efficiency is approximately given by the
average of these bounds and reads
\begin{equation}
  \label{Tmax}
  T^\mathrm{max}=\frac{E_W}{\ln(\nu/f)}.
\end{equation}

\subsection{Binary system}
\label{sec:binary}

The binary system is a particular case of the model described in
\sref{sec:definition-model}, where the distribution of binding energies
is discrete and takes only two values $E_{W_1}$ or $E_{W_2}$, with
$E_{W_1}<E_{W_2}$~\cite{wolff10-react-two-dimen-bin-diso}.  The
corresponding $S_1$ and $S_2$ sites are labeled ``shallow'' and ``deep''
for the lower and the higher binding energy, respectively, and
subscripts only refer to the \emph{type} of site $i$ here ($i=1,2$).
In general, discreteness of particles and fluctuations in their number
may render the mean-field description of the stochastic particle system at
hand unsuitable, and thus rate equations overestimate the
efficiency~\cite{tielens95, biham02, krug03, biham05, lohmar06,
  lederhendler08}.  For binary systems of sufficient size as treated
here, these effects are negligible, and one finds excellent agreement
between the efficiency seen in KMC simulations and obtained from a rate
equation description~\cite{wolff10-react-two-dimen-bin-diso}.
The steady-state equations of this model for the number of particles
$N_i$ on sites of type $i$ read
\begin{equation}
  \label{rate-eqs}
  \begin{aligned}
    \frac{\dd N_1}{\dd t} &=
    f(S_1-N_1) -W_1N_1
    -A_1N_1(S_2-N_2)   
    -A_1N_1N_2         
    \\ &\quad
    -2A_1N_1^2         
    +A_2N_2(S_1-N_1)   
    -A_2N_1N_2         
    = 0, \\
    \frac{\dd N_2}{\dd t} &=
    f(S_2-N_2) -W_2N_2
    -A_2N_2(S_1-N_1) -A_2N_1N_2
    \\ &\quad
    -2A_2N_2^2
    +A_1N_1(S_2-N_2) -A_1N_1N_2
    = 0,
  \end{aligned}
\end{equation}
where the \emph{sweeping rate} $A_i = a_i/S$ governs the rate of
reaction due to hops from sites of type $i$ (not an elementary process).
The simple expression for $A_i$ results from the fact that for
disordered systems, a consistent rate equation description has to assume
that particles can hop from any site to any
other~\cite{wolff10-react-two-dimen-bin-diso}.
Note that this is not mandatory for \emph{homogeneous} systems, where
using $A=a/S$ to describe nearest-neighbor hopping presents an
unsubstantiated approximation~\cite{lohmar06, lohmar08,
  lohmar09-accur-rate-coeff}.
While~\eref{rate-eqs} are exactly solvable by finding the real positive
root of a third-order polynomial, the result is cumbersome, hence we
prefer a direct numerical solution throughout.

The reaction terms provide the production rate of the process.
Adding up all terms proportional to the $A_i$ in $\dd N/\dd t =\dd
N_1/\dd t +\dd N_2/\dd t$, mere hopping terms (not leading to a
reaction) cancel.  Since the reaction consumes two particles, we obtain
the rate at which particles are removed by the reaction as
\begin{equation}
  \label{2R}
  2R = 2A_1N_1^2+2A_2N_2^2+2(A_1+A_2)N_1N_2, 
\end{equation}
where the three contributions correspond to reactions triggered by
hopping between shallow sites, between deep sites, and from one type of
site to the other, respectively.
Relating this to the total particle influx $f(S_1+S_2)=fS$ yields the
\emph{efficiency} $\eta = 2R/(fS)$.

In this system, whenever there is a substantial efficiency of
reaction, the two types of sites play entirely different roles:
Shallow sites provide \emph{mobility} to the particles---by allowing
them to move easily and quickly, they can sweep large parts of the
system.  Deep sites, however, provide \emph{coverage}---while particles
stuck there hardly ever move, they are also prevented from leaving the
system.  Acting together, the shallow sites allow particles to traverse
over many sites of the system, funneling them to deep sites eventually,
where they likely meet a stuck particle to react with and contribute to
the efficiency.  As a result, deep sites are, on average, covered half
of the time, since occupying an empty site takes as long as emptying an
occupied one by a reaction.

Compared with this process, reactions taking place on shallow sites are
very unlikely.  As soon as particles on such sites become mobile enough,
they may sweep over shallow sites only, which are sparsely populated
compared with deep sites.  Hence they most likely end up in a deep well
which is already occupied (leading to a reaction), or the well is newly
occupied by the incoming particle, now stuck and waiting for the next
one.  Likewise, reaction by hops between deep sites is prevented
since particles only become mobile at very high temperatures, when the
overall coverage (governed by $f/W$) is low, and traversal over shallow
wells is a leak through which particles leave immediately.

\begin{figure}
  \centering
  \tikzsetnextfilename{fig3}%
  \begin{tikzpicture}
    \begin{axis}[
      xlabel=Temperature $T/\mathrm K$
      ,ylabel={\color{blue}Cvg.\ $\theta$, \color{red}Eff.\ $\eta$}
      ,width=\singleplotwidth
      ,xmin=8,xmax=38,ymin=0,ymax=1.05
      ,minor tick num=4
      ,x tick label style={/pgf/number format/precision=0}
      ,y tick label style={/pgf/number format/precision=1}
      ]
      \pgfplotstableread{NdST-binre-750longhop-25.dat}\tableNdS
      \pgfplotstableread{etaT-binre-750longhop-25.dat}\tableeta
      \addplot[supp graph,blue,densely dotted] plot
      table[x index=0,y index=2] from \tableNdS;
      \addplot[supp graph,blue,densely dashed] plot
      table[x index=0,y index=3] from \tableNdS;
      \addplot[main graph,blue,solid] plot table[x index=0,y index=1] from
      \tableNdS;
      \addplot[supp graph,red,densely dotted] plot
      table[x index=0,y index=2] from \tableeta;
      \addplot[supp graph,red,densely dashed] plot
      table[x index=0,y index=3] from \tableeta;
      \addplot[supp graph,red,mydashdot] plot
      table[x index=0,y index=4] from \tableeta;
      \addplot[main graph,red,solid] plot table[x index=0,y index=1] from
      \tableeta;
    \end{axis}
  \end{tikzpicture}
  \caption{Coverage $\theta$ (blue) and efficiency $\eta$ (red) in the
    binary model (obtained by the rate equation approach) as functions
    of temperature $T$.  Partial and total coverages (all in blue)
    $N_1/S_1$ (dotted), $N_2/S_2$ (dashed), $\theta=N/S$ (solid).
    Efficiency contributed (see~\eref{2R}, all in red) by events on
    shallow sites (dotted), on deep sites (dashed), and by hopping from
    one to the other type (dash-dotted), as well as total efficiency
    $\eta$ (solid).  Unlike partial coverages, efficiency contributions
    add up to the total efficiency.  Parameters chosen as described in
    \sref{sec:kmc}, using mean activation energies for
    standard\,/\,shallow sites ($S_1/S=75\%$), and activation energies
    for the deep sites ($S_2/S=25\%$) enhanced by
    $E_{W_2}-E_{W_1}=750$~K.  For intermediate temperatures, the
    efficiency is dominated by reaction events due to hops from shallow
    to deep sites.}
  \label{fig:cvgetabin}
\end{figure}
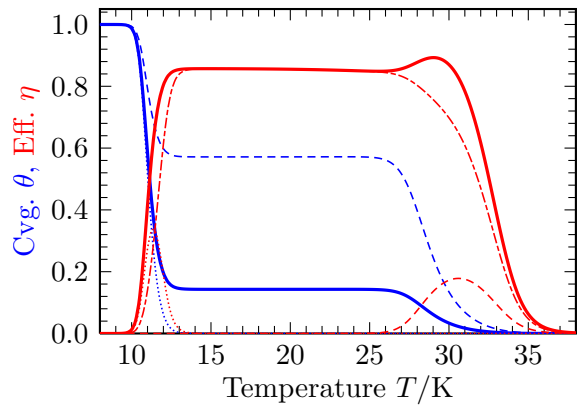

The different role that shallow and deep sites play in this model is
highlighted by \fref{fig:cvgetabin}, obtained by
solving~\eref{rate-eqs}.  For a wide range of parameters, the separate
efficiency windows corresponding to \emph{homogeneous} systems of either
type of site are ``bridged'' to one broad efficiency window, between the
lower temperature bound of the shallow sites and the upper temperature
bound of the deep sites.  In stark contrast to the homogeneous case, the
coverage over this whole temperature range is approximately constant.

\section{Continuous case and effective model}
\label{sec:effective}

Consider now a particular realization of the continuous-distribution
model.  We propose a mapping of such a system to a binary-distribution
system that reproduces the efficiency (as well as the coverage) found in
simulations to surprising accuracy.  We want to emphasize that this
mapping is an entirely analytical prescription, and that it does not
involve any data obtained from simulations of the system.

\subsection{The effective binary system}
\label{sec:mapping-idea}

The central idea of our mapping is that an energy landscape drawn from a
continuous distribution of binding energies can be condensed to only two
types of sites: The effectively ``shallow'' sites, which have low
binding energy, and which provide particles with easy mobility and
funnel them, namely into the effectively ``deep'' sites, which have high
binding energy and which provide sufficient coverage.
If this partition into shallow and deep sites is performed at the proper
energy $E_\mathrm{cut}$, and if the binding energies of the two
effective types of sites are chosen appropriately, the original detailed
binding energy of each individual site in a realization of the
continuous case is irrelevant.  Shallow sites in the effective model
will reflect the overall mobility of particles on sites with energies
smaller than $E_\mathrm{cut}$.  Deep sites in the effective model will
capture the overall ability to bind particles strong enough to provide
coverage on sites with energies larger than $E_\mathrm{cut}$.
Since both diffusion and desorption are thermally activated processes,
it is obvious already at this stage that the threshold $E_\mathrm{cut}$
must increase with temperature.

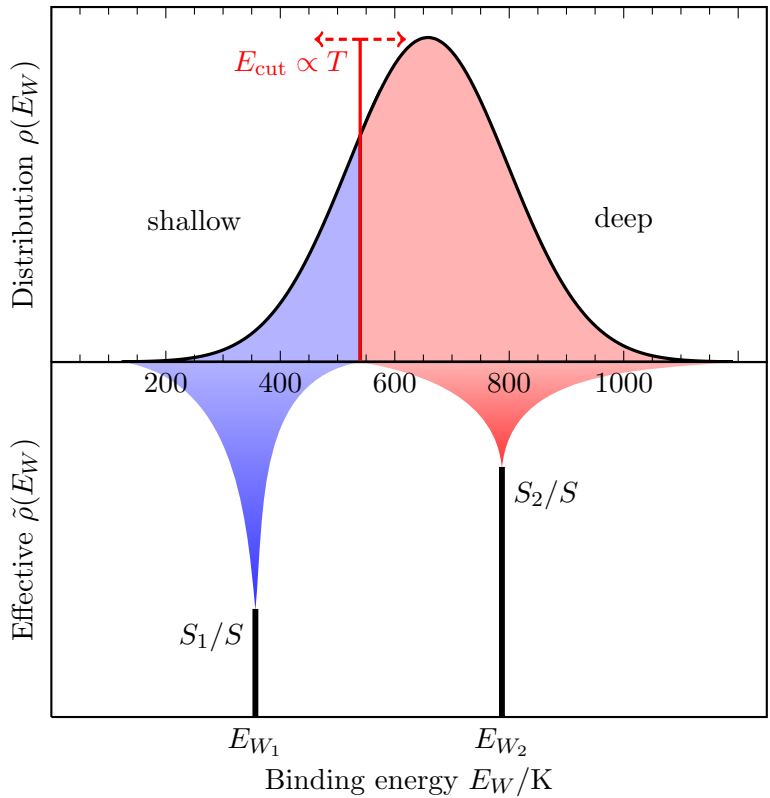
\begin{figure}
  \centering
  \tikzsetnextfilename{fig4}%
  \begin{tikzpicture}[relative=false]
    \pgfmathsetmacro{\mean}{658}\pgfmathsetmacro{\stdev}{0.3*\mean}
    \pgfmathsetmacro{\Ecut}{\mean-0.6*\stdev}
    \pgfmathsetmacro{\fs}{0.3}\pgfmathsetmacro{\fd}{0.7}
    \pgfmathsetmacro{\Es}{\mean-sqrt(\fd/\fs)*\stdev}
    \pgfmathsetmacro{\Ed}{\mean+sqrt(\fs/\fd)*\stdev}
    \pgfmathsetmacro{\Emin}{\mean-2.7*\stdev}
    \pgfmathsetmacro{\Emax}{\mean+2.7*\stdev}
    \pgfmathsetmacro{\pref}{1/(sqrt(2*pi)*\stdev)}
    \begin{axis}[name=binary,scale only axis 
      ,width=0.6\linewidth,height=0.3\linewidth
      ,xlabel=Binding energy $E_W/\mathrm K$
      ,ylabel=Effective $\tilde\rho(E_W)$
      ,xmin=0,xmax=1250,ymin=0,ymax=1.0
      ,xtick={\Es,\Ed},xticklabels={$E_{W_1}$,$E_{W_2}$}
      ,ytick={\empty}
      ]
      \addplot[main graph
      ,ybar,bar width=1pt] plot coordinates {(\Es,\fs) (\Ed,\fd)};
      \draw (axis cs:\Es,\fs) node[below left] {$S_1/S$};
      \draw (axis cs:\Ed,\fd) node[below right] {$S_2/S$};
      \shade[top color=blue!30!white,bottom color=blue!80!white]
      (axis cs:\Emin,1.0) to[out=-10,in=95] (axis cs:\Es,\fs)
        to[out=85,in=190] (axis cs:\Ecut,1.0)
        -- cycle;
      \shade[top color=red!30!white,bottom color=red!80!white]
      (axis cs:\Ecut,1.0) to[out=-10,in=95] (axis cs:\Ed,\fd) 
        to[out=85,in=185] (axis cs:\Emax,1.0) -- cycle;
    \end{axis}
    \begin{axis}[scale only axis,at={(binary.north)},anchor=south
      ,width=0.6\linewidth,height=0.3\linewidth
      ,ylabel=Distribution $\rho(E_W)$
      ,xmin=0,xmax=1250,ymin=0,ymax=\pref
      ,enlarge y limits=upper
      ,xtick={0,200,400,...,1200},xticklabels={,200,400,...,1000}
      ,ytick={\empty}
      ,minor tick num=3
      ,axis on top=true 
      ]
      \addplot[main graph,fill=blue!30!white
      ,samples=50,domain={\Emin}:{\Ecut+30}]
      plot expression {exp(-1*((x-\mean)/\stdev)^2)/(sqrt(2*pi)*\stdev)}
      \closedcycle;
      \addplot[main graph,fill=red!30!white
      ,samples=50,domain={\Ecut}:{\Emax}]
      plot expression {exp(-1*((x-\mean)/\stdev)^2)/(sqrt(2*pi)*\stdev)}
      \closedcycle;
      \draw[plot level,red] (axis cs:\Ecut,0) -- (axis cs:\Ecut,\pref)
      node[below left] {$E_\mathrm{cut} \propto T$};
      \draw[->,densely dashed,very thick,red]
      (axis cs:\Ecut,\pref) -- +(axis cs:-80,0);
      \draw[->,densely dashed,very thick,red]
      (axis cs:\Ecut,\pref) -- +(axis cs:+80,0);
      \draw (rel axis cs:0.20,0.4) node {shallow};
      \draw (rel axis cs:0.80,0.4) node {deep};
      \end{axis}
  \end{tikzpicture}
  \caption{Basic idea of the mapping to the effective binary model.}
  \label{fig:mapping}
\end{figure}

\Fref{fig:mapping} depicts this central idea, the notation and further
details will be described in the next section.  Note that we always
consider the effective binary system to be well-mixed---it is essential
that deep sites are as easily accessible as possible from the shallow
sites.  This also ensures that the system is well described by rate
equations, which we will employ below.
Moreover, we will focus on the case that there are still a reasonable
number of both types of sites, or equivalently, that the continuous
distribution is still sampled well on both sides of the threshold energy
$E_\mathrm{cut}$.  The issue of rare events and sample-to-sample
fluctuations will be returned to in \sref{sec:realizations}.  In the
relevant case that these fluctuations are sufficiently small, our
mapping is equivalent to a mapping from the entire continuous
\emph{distribution} of binding energies to a binary one.  Unless
specified otherwise, we will always refer to the latter mapping in the
remainder.

\subsection{Confirmation of mapping assumptions by simulations}
\label{sec:mapping-check}

The idea presented above has to be tested before we progress.  To this
end, we first introduce some additional notation and define the relevant
quantities.
We use subscripts $\omega$ for a realization of binding energies and $s$
for a single site.  Then $E_{\omega,s}$ is the binding energy of site
$s$ in realization $\omega$ (in this subsection, we will omit the
subscript $W$ for brevity).
Further, $r_{\omega,s}$ denotes the steady-state (or time-averaged)
fraction of all reaction events in realization $\omega$ that takes place
on site $s$; we call $r_{\omega,s} \in[0,1]$ the \emph{reactivity}.
We denote by $u_{\omega,s}$ the steady-state fraction of (physical) time
that site $s$ in realization $\omega$ is occupied; $u_{\omega,s}
\in[0,1]$ is called the \emph{occupancy}.

We are interested in the relation between the binding energy of a
certain site and its reactivity.  On that account, we transform from the
spatial distribution of reaction events to the distribution with respect
to the local energy,
\begin{equation}
  r_\omega(E) = \sum_s r_{\omega,s} \delta(E - E_{\omega,s}).
\end{equation}
Obtained from a limited number of sites, $r_\omega(E)$ is obviously only
a collection of $S$ sample values from an imagined smoothed function.
Gathering information from a set $\{\omega\}$ of realizations, we
additionally have to weight this distribution for each single
realization $\omega$ according to the efficiency $\eta_\omega$ of the
latter, effectively accounting for the \emph{number} of reaction events
during a certain period of time: A site with given energy might be
responsible for a much larger fraction in one realization simply because
in distant parts of the surface the particular energy landscape results
in fewer events.  In this case the overall efficiency of this particular
realization will be diminished, and rescaling by the efficiency removes
this unwanted distortion.  The result (which we still call
``reactivity'') reads
\begin{equation}
  r_{\{\omega\}}(E) =
  \frac{\sum_\omega r_{\omega}(E) \cdot \eta_\omega}{\sum_\omega \eta_\omega},
\end{equation}
including proper normalization
\begin{equation}
  \sum_E r_{\{\omega\}}(E) =
  \frac{\sum_\omega \eta_\omega \sum_s r_{\omega,s}
    \sum_E \delta(E -E_{\omega,s})}
  {\sum_\omega \eta_\omega} =
  \frac{\sum_\omega \eta_\omega \left(\sum_s r_{\omega,s}\right)}
  {\sum_\omega \eta_\omega}
  = 1.
\end{equation}
The slightly clumsy notation is an artifact of the finite number of
samples.  In the limit considering the statistical ensemble of
\emph{all} possible realizations, the $E$ sum becomes an integral, and
functions of $E$ become smooth.

For the occupancy of sites of a certain energy, we transform
analogously to the above.  Comprising several realizations does not need
any weighting here, since the definition of the occupancy
$u_{\omega,s}$ of a site does not relate to the total coverage in the
realization.  Normalization, however, implies we factor out the total
number of particles in all realizations.  Noting that the time-averaged
total coverage in realization $\omega$ reads $N_\omega = \sum_s
u_{\omega,s}$, we have
\begin{equation}
  u_{\{\omega\}}(E) =
  \frac{\sum_\omega u_{\omega}(E)}{\sum_\omega N_\omega},
\end{equation}
again called ``occupancy''.
Then
\begin{equation}
  \sum_E u_{\{\omega\}}(E) =
  \frac{\sum_\omega \sum_s u_{\omega,s}
    \sum_E \delta(E -E_{\omega,s})}
  {\sum_\omega  N_\omega} =
  \frac{\sum_\omega \left(\sum_s u_{\omega,s}\right)}
  {\sum_\omega N_\omega}
  = 1.
\end{equation}

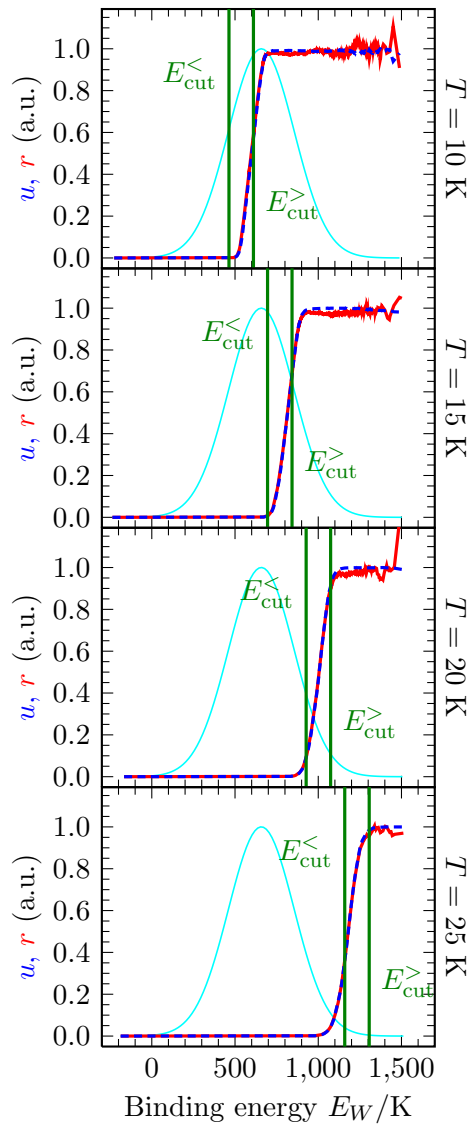
\begin{figure}
  \centering
  \tikzsetnextfilename{fig5}%
  \begin{tikzpicture}
    \begin{groupplot}[
      group style={
        ,group size=1 by 4
        ,glued}
      ,width=\stepplotwidth,height=\stepplotheight
      ,xlabel=Binding energy $E_W/\mathrm K$
      ,ylabel={\color{blue}$u$, \color{red}$r$ \color{black}(a.u.)},
      ,xmin=-300,xmax=1700
      ,ymin=-0.05,ymax=1.19
      ,xprec=0
      ,yprec=1
      ,minor x tick num=4,minor y tick num=3
      ]
      \pgfplotsinvokeforeach{10,15,...,25} {
        \nextgroupplot
        \extplot[id={nrm 30 #1}] ext-plots.py;
        \Ecuts{#1}
        \draw[plot level,darkgreen] (axis cs:\Ecuthi,-0.05)
        -- (axis cs:\Ecuthi,1.19)
        node[near end,left] {\color{darkgreen}$E_\mathrm{cut}^<$};
        \draw[plot level,darkgreen] (axis cs:\Ecutlo,-0.05)
        -- (axis cs:\Ecutlo,1.19)
        node[near start,right] {\color{darkgreen}$E_\mathrm{cut}^>$};
      }
    \end{groupplot}
    \foreach \T [count=\row] in {10,15,...,25}
      \node[tight,rotate=-90,anchor=south] at (group c1r\row.east) 
      {$T = \T\ \mathrm{K}$};
  \end{tikzpicture}
  \caption{(Rescaled) occupancy $u(E_W)$ (blue, dashed) and reactivity
    $r(E_W)$ (red) versus binding energy $E_W$ of the sites, for the
    normal distribution of relative width $\tilde\sigma = 30\%$ and at
    temperatures $T=10,\,15,\,\dots,\, 25\ \mathrm K$ (top to bottom).
    The rescaled PDF (thin cyan line) is shown for reference.  The
    vertical green lines mark two specific choices for $E_\mathrm{cut}$
    to be defined in
    \sref{sec:mapping-derivation}.}\label{fig:steps-of-E}
\end{figure}

\Fref{fig:steps-of-E} shows the occupancy and the reactivity as
functions of the binding energy, comprised from KMC simulations for $10$
realizations.
We have chosen the paradigmatic example of the normal distribution here,
with a relative width of $\tilde\sigma = 30\%$, and for several
temperatures.
The unprocessed functions $u_{\{\omega\}}(E)$ and $r_{\{\omega\}}(E)$
are \emph{not} shown, as they exhibit strong fluctuations (we return to
this issue in \sref{sec:realizations}).  Instead, we present better
approximations of the smooth ensemble averages, which we denote by
$u(E)$ and $r(E)$.  These approximations are obtained by a sliding
average, in which the function's value at each sample energy $E$ is
replaced by the average of all points within a certain energy
neighborhood.  This is preferable to an average over a fixed number of
neighboring points, since samples are not equally spaced on the energy
scale.  In our plots we use an energy interval of $4\%$ of the total
range of energies sampled.
The original distribution of binding energies is drawn as a thin line
for orientation.  In order that the plots can be easily compared, we
have rescaled this distribution to have a maximum value of unity.
Likewise, we have rescaled $u(E)$ and $r(E)$ by a constant factor such
that $u(E)$ attends a maximal value of unity as well.

Whenever sample-to-sample fluctuations become small, both occupancy and
reactivity very clearly distinguish two types of sites according to
their energy, with a fairly steep ``step'' between them.
For lower energies there is little to no activity, and we identify these
sites as effectively shallow.  The sites with higher energies, however,
provide nearly all the coverage and reaction events, and are effectively
deep.  Moreover on both sides of this border, the precise energy of the
individual sites does no longer play any role.

\begin{figure}
  \centering
  \tikzsetnextfilename{fig6a}%
  \begin{tikzpicture}
    \begin{axis}[
      ,width=\stepplotwidth,height=\stepplotheight
      ,xlabel=Binding energy $E_W/\mathrm K$
      ,ylabel={\color{blue}$u$, \color{red}$r$ \color{black}(a.u.)}
      ,xmin=400,xmax=2600
      ,ymin=-0.05,ymax=1.19
      ,xprec=0
      ,yprec=1
      ,minor x tick num=4,minor y tick num=3
      ]
      \extplot[id={exp 30 30}] ext-plots.py;
      \Ecuts{30}
      \draw[plot level,darkgreen] (axis cs:\Ecuthi,-0.05)
      -- (axis cs:\Ecuthi,1.19)
      node[near end,left] {\color{darkgreen}$E_\mathrm{cut}^<$};
      \draw[plot level,darkgreen] (axis cs:\Ecutlo,-0.05)
      -- (axis cs:\Ecutlo,1.19)
      node[near start,right] {\color{darkgreen}$E_\mathrm{cut}^>$};
    \end{axis}
  \end{tikzpicture}\qquad%
  \tikzsetnextfilename{fig6b}%
  \begin{tikzpicture}
    \begin{axis}[
      ,width=\stepplotwidth,height=\stepplotheight
      ,xlabel=Binding energy $E_W/\mathrm K$
      ,ylabel={\color{blue}$u$, \color{red}$r$ \color{black}(a.u.)}
      ,xmin=150,xmax=1150
      ,ymin=-0.05,ymax=1.19
      ,xprec=0
      ,yprec=1
      ,minor x tick num=3,minor y tick num=3
      ]
      \extplot[id={uni 40 10}] ext-plots.py;
      \Ecuts{10}
      \draw[plot level,darkgreen] (axis cs:\Ecuthi,-0.05)
      -- (axis cs:\Ecuthi,1.19)
      node[near end,left] {\color{darkgreen}$E_\mathrm{cut}^<$};
      \draw[plot level,darkgreen] (axis cs:\Ecutlo,-0.05)
      -- (axis cs:\Ecutlo,1.19)
      node[near start,right] {\color{darkgreen}$E_\mathrm{cut}^>$};
    \end{axis}
  \end{tikzpicture}
  \caption{As in \fref{fig:steps-of-E}, $u(E_W)$ (blue, dashed) and
    $r(E_W)$ (red), for an exponential distribution with $\tilde \sigma
    = 30\%$ at $T = 30\ \mathrm K$ (left), and for a uniform
    distribution with $\tilde\sigma = 40\%$ at $T = 10\ \mathrm K$
    (right).  Rescaled PDF's (thin cyan line) shown for
    reference.}\label{fig:steps-of-E2}
\end{figure}
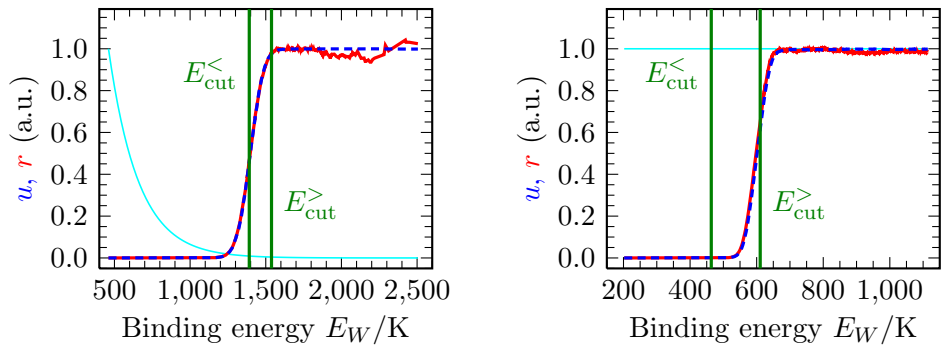

The threshold energy $E_\mathrm{cut}$ separating both types of sites
moves to higher energies as the temperature increases, as expected.  For
very high temperatures, only very few sites from the distribution tail
still contribute coverage and reactions.  On the other hand, we checked
that the cut is largely independent of the shape and width of the
distribution.  Again, this is compatible with our earlier thoughts and
will be substantiated in the next subsection.
\Fref{fig:steps-of-E2} gives one example for the exponential and the
uniform distribution, respectively, qualitatively similar to the results
for the normal distribution.  For the exponential distribution,
high-energy tails decay more slowly than for the normal one, leading to
a picture for $T=30\ \mathrm K$ which resembles those for lower
temperatures in the normal-distribution case (with deep sites still over
a large range of energies).  The plot for the uniform distribution at
$T=10\ \mathrm K$ hardly shows fluctuations in the (smoothed) occupancy
and reactivity, compared with the normal distribution.  This is due to the
fact that there are no high-energy tails, so that energies are roughly
equally (and ``densely'') spaced up to their maximum, and hence few
outliers do not affect the smoothed plot at all.

Lastly, we find the agreement between the graphs for occupancy and
reactivity remarkable in all instances.  This shows that exactly the
highly-occupied sites are those on which reactions take place, just as
in the genuinely binary model.  Together we thus have numerical proof of
the arguments we have presented in \sref{sec:mapping-idea}.

\subsection{Heuristic derivation of the mapping}
\label{sec:mapping-derivation}

There are a multitude of strategies to arrive at a mapping from the
continuous to the binary model.  We should state very clearly that, in
fact, one could even (implicitly) define an effective \emph{homogeneous}
model which uses only \emph{one} type of site.  Since a homogeneous
system can exhibit the whole range of values $[0,1]$ of the efficiency
$\eta$, it is mathematically trivial that, if we may choose \emph{all}
parameters appropriately, it can reproduce the efficiency found in the
continuous model.  Apart from the implicitness of such a definition,
this does not lead to any understanding of the underlying physics,
however.

Our approach is different.  We try to retain as many features of the
continuous-distribution model as possible, altering only a minimal
subset of parameters---this way, we minimize the arbitrariness of the
mapping.
The physically plausible strategy is to restrict changes to the binding
energy distribution itself; we will find an equivalent system where only
the ``well depths'' of sites are changed (to a binary distribution).
Concretely, this implies that we keep $f$, $T$, $S$ and $\nu$ fixed.

Note that we speak of a mapping of the binding energies $E_W$
throughout, but the hopping barriers $E_a$ are of similar importance.
In the continuous case they are fixed by demanding constant $\Delta E =
E_W - E_a$ for all sites, which guarantees detailed balance, and implies
a constant ratio $W/a$.
We stick to this choice (particularly for the detailed-balance
argument), using the same $\Delta E$ for the binary distribution.

Consequently, there are three quantities left that define the mapping:
the discrete binding energies $E_{W_1}$ and $E_{W_2}$ of shallow and
deep sites in the effective model, respectively, and the number of both
types of sites, conveniently parametrized by a \emph{cutting energy}
$E_\mathrm{cut}$ at which we split the distribution of binding energies.

The discrete binary distribution has to capture the essential features
of the original continuous one.  The two most obvious properties are the
mean binding energy $\langle E_W \rangle$ and the standard deviation
$\sigma$.  We demand that the effective binary distribution has the same
mean and standard deviation.\footnote{We also tested several other
  choices to set $E_{W_1}$ and $E_{W_2}$ once $S_1$ and $S_2$ are known.
  They all introduce additional arbitrariness without improving (but
  often detracting from) the quality of the mapped model predictions.}
Given the number of shallow and deep sites, $S_1$ and $S_2$, this fixes
the discrete binding energies to read
\begin{equation}\label{E-W-eff}
  E_{W_1} = \langle E_W \rangle - \sigma \sqrt{S_2/S_1},
  \qquad
  E_{W_2} = \langle E_W \rangle + \sigma \sqrt{S_1/S_2}.
\end{equation}
Recall our earlier assumption that both $S_1$ and $S_2$ are not too
small, and in particular that the binary model does not degenerate to a
homogeneous model in the regime of our interest.

It only remains to choose $E_\mathrm{cut}$, which governs how many sites
are regarded as shallow and deep.  As the simplest choice, we assume
that this energy is independent of the shape and parameters of the
distribution.

Now consider the limiting regime of high temperature.  There are many
shallow and few deep sites, regardless of the precise form of
$E_\mathrm{cut}$.  The limiting factor for the efficiency is lack of
coverage on the deep sites, while mobility on shallow sites (to quickly
funnel atoms to the deep ones) is a given.  Hence, we have to split the
binding energy distribution at an $E_\mathrm{cut}$ such that at this and
at higher energies, sites are sufficiently occupied.  This energy is set
by $f = \eval{W}_{E_W=E_\mathrm{cut}}$, such that without any reactions,
we would have half-filling on average.  This means
\begin{equation}
  E_\mathrm{cut} = E_\mathrm{cut}^< :=
  T \ln(\nu/f) \quad \text{(high $T$)}.
\end{equation}
Mobility on the shallow sites is then guaranteed by $\eval{a}_{E_W \leq
  E_\mathrm{cut}} \geq \eval{a}_{E_W=E_\mathrm{cut}} \gg
\eval{W}_{E_W=E_\mathrm{cut}} = f$.

For low temperatures, on the other hand, the overall coverage is high,
and there are few shallow but lots of deep sites.  In this regime the
efficiency is not limited by lack of coverage, but by a lack of mobility
on the shallow sites.  Particles have to be able to hop at least as
frequently as new ones arrive, or else LH rejection will curtail the
efficiency.
Therefore, the maximal energy of ``working'' shallow sites is set by $f
= \eval{a}_{E_W=E_\mathrm{cut}}$, and sites at lower energies have $a >
f$.  Re-writing the condition using $E_a = E_W - \Delta E$ we obtain
\begin{equation}
  E_\mathrm{cut} = E_\mathrm{cut}^> :=
  T \ln(\nu/f) + \Delta E \quad \text{(low $T$)}.
\end{equation}
On the deep sites, we then have $f \geq \eval{a}_{E_W \geq
  E_\mathrm{cut}} \gg W$, so high coverage there is guaranteed.

These choices for $E_\mathrm{cut}$ are shown as vertical lines in the
energy-resolved pictures for occupancy and reactivity,
\fref{fig:steps-of-E} and \fref{fig:steps-of-E2}.  On the other hand,
the latter quantities suggest a reasonable choice for $E_\mathrm{cut}$
themselves, say, the energy at which occupancy and reactivity reach half
of their maximal value.  Comparing these choices, $E_\mathrm{cut}$ as
suggested by the plots is always found between $E_\mathrm{cut}^<$ and
$E_\mathrm{cut}^>$ (as defined above).  For low temperatures the
suggested cutting energy indeed moves closer to the upper energy
$E_\mathrm{cut}^>$, whereas with increasing temperature it comes ever
closer to the lower energy $E_\mathrm{cut}^<$.  Our heuristic derivation
of the threshold energy is hence confirmed by numerical simulations.

The shift of $\Delta E$ between the two choices reflects the fact that
in the above arguments, for high temperatures the distribution of
\emph{binding energies} is cut in two, while at low temperatures the
distribution of \emph{diffusion barriers} is cut.  These results are
indeed independent of the shape, mean, and width of the distribution, as
suggested above.
The limiting temperature regimes correspond to finite temperatures (not
to $T\to 0$ and $T\to\infty$, that is), hence the transition between
them involves additional temperature scales (describing location and
width).  These quantities evidently have to depend on the shape and
width of the distribution, which we will confirm by simulation results
(see below).  We have not found a convincing theoretical argument to
determine these scales.
In practice, it is still straightforward to determine the appropriate
choice for $E_\mathrm{cut}$.  Sites with energies in the range
$[E_\mathrm{cut}^<, E_\mathrm{cut}^>]$ provide coverage \emph{and}
mobility, and could be labeled \emph{either} deep or shallow, depending
on this choice.  The proper choice of $E_\mathrm{cut}$ in the limits of
low and high temperature regards these sites to provide the scarce
property which limits the efficiency, respectively.  The ``opposite''
choice misinterprets their role and leads to substantially lower
efficiency.\footnote{For narrow exponential and uniform distributions
  ($\tilde \sigma = 10\%$), this statement is not strictly true: There
  is a very small range of low temperatures at which the ``high-$T$''
  model has \emph{marginally higher} efficiency, starting where it is
  still degenerate (no shallow sites) and ending just after it features
  both effectively shallow \emph{and} deep sites.  This is an artifact
  of the simple mapping prescription, and in any case, such narrow
  distributions are not the focus of this work.}  Summing up, in both
limits, one chooses $E_\mathrm{cut}$ such that it leads to the
\emph{highest possible efficiency in the effective model}.  It is then
plausible to stick with this prescription for intermediate temperatures
as well.

In principle, cutting the distribution at $E_\mathrm{cut}$ provides real
values for the numbers $S_i$ of sites of a given type.  To stay true to
the idea that we replace the whole energy landscape by an effective one,
we round to the nearest integer values for $S_i$, when the effective
system has physically sensible parameters throughout.  In the case where
the system is sufficiently large and still has a substantial number of
both shallow as well as deep sites, the difference to the nearest
integer values is negligible anyway.

\subsection{Comparison to KMC simulations}
\label{sec:comparison}

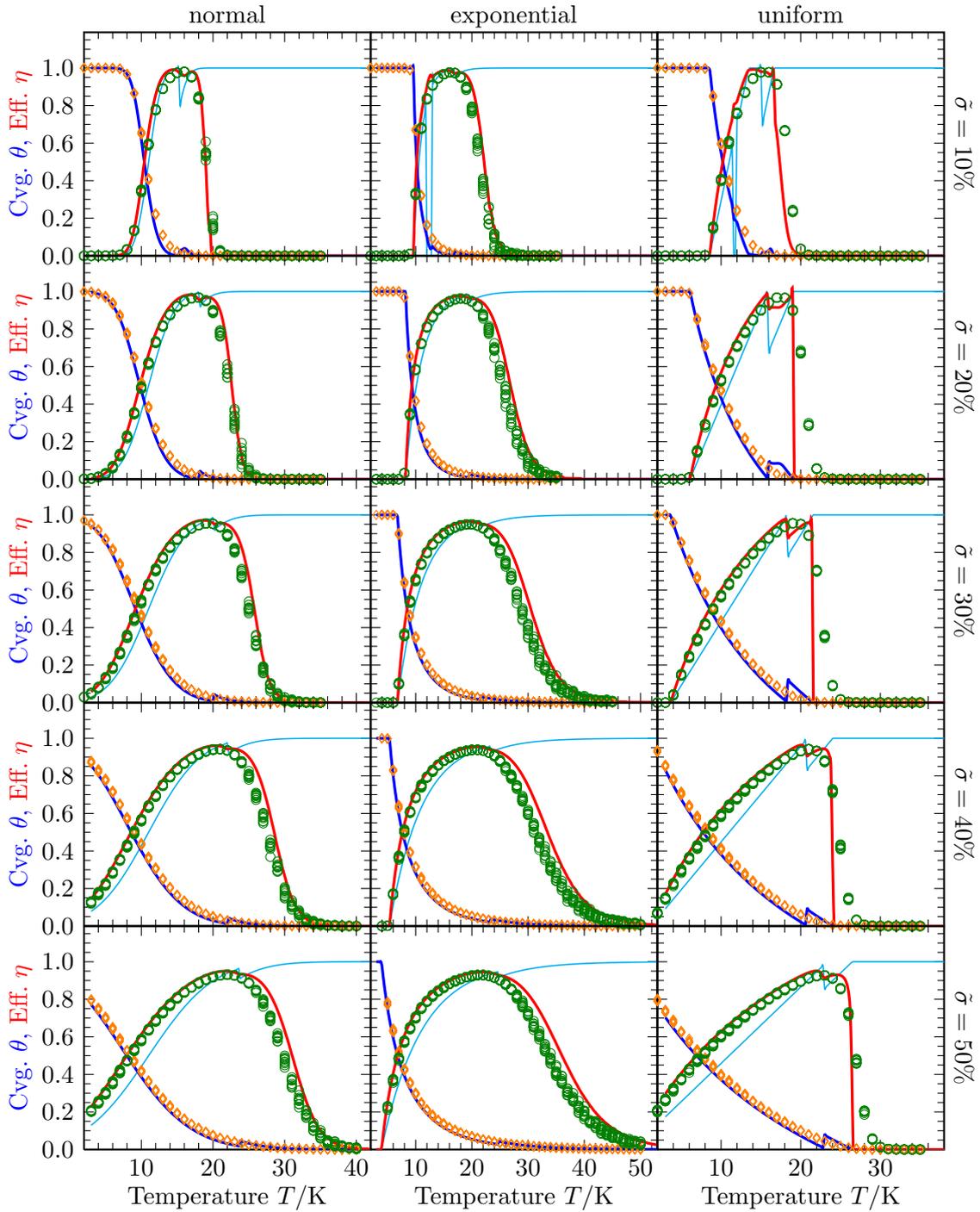
\begin{figure}
  \centering
  \tikzsetnextfilename{fig7}%
  \begin{tikzpicture}
    \begin{groupplot}[
      group style={
        group size=3 by 5,
        glued}
      ,width=\stepplotwidth,height=\stepplotheight
      ,xlabel=Temperature $T/\mathrm K$
      ,ylabel={\color{blue}Cvg.\ $\theta$, \color{red}Eff.\ $\eta$}
      ,xmin=2,ymin=0,ymax=1.19
      ,xprec=0
      ,yprec=1
      ,minor x tick num=4,minor y tick num=3
      ,every axis title/.append style={text depth=0pt}  
      ]
      \def\totalcolumns{3}\def\totalrows{5}
      \pgfmathtruncatemacro\totalplots{\totalcolumns*\totalrows}
      \pgfplotsinvokeforeach{1,2,...,\totalplots} {
        \pgfmathtruncatemacro\currentrow{(#1-0.1)/\totalcolumns+1}
        \pgfmathtruncatemacro\currentcolumn{#1-\totalcolumns*(\currentrow-1)+0.1}
        \ifcase\currentcolumn  
          \or \def\dst{nrm} \def\txt{normal} \def\xmax{42} 
          \or \def\dst{exp} \def\txt{exponential} \def\xmax{53} 
          \or \def\dst{uni} \def\txt{uniform} \def\xmax{38} 
        \fi
        \pgfmathtruncatemacro\tspct{10*\currentrow}
        \ifnum\currentrow=1\def\atitle{\txt}\else\def\atitle{}\fi
        \edef\nextcmds{
          \noexpand\nextgroupplot[xmax=\xmax,title=\atitle]
          \noexpand\pgfplotstableread{../10-diso/TetaN-\dst-0\tspct.dat}\noexpand\KMC
          \noexpand\pgfplotstableread{effbin-maxeta-\dst-\tspct.dat}\noexpand\EFF
          \noexpand\addplot[supp graph,cyan] plot table[x index=0,y index=3]
          from \noexpand\EFF;
          \noexpand\addplot[main graph,blue] plot table[x index=0,y index=1]
          from \noexpand\EFF;
          \noexpand\addplot[mark=diamond,orange]
          plot table[x index=0,y expr=\noexpand\thisrowno{2}/10000] from \noexpand\KMC;
          \noexpand\addplot[main graph,red] plot table[x index=0,y index=2] 
          from \noexpand\EFF;
          \noexpand\addplot[mark=o,darkgreen] plot table[x index=0,y index=1]
          from \noexpand\KMC;
        }\nextcmds
      }
    \end{groupplot}
    \foreach \tspct [count=\row] in {10,20,...,50}
    \node[tight,rotate=-90,anchor=south] at (group c3r\row.east) 
    {$\tilde\sigma = \tspct\%$};
  \end{tikzpicture}
  \caption{Coverage $\theta$ (orange diamonds, blue lines) and
    efficiency $\eta$ (green circles, red lines) versus temperature $T$,
    of the continuous-distribution system from KMC simulations (marks,
    one per realization), and as obtained for the effective binary model
    via rate equations (lines).  Also shown is the fraction $S_1/S$ of
    shallow sites in the effective model (thin cyan line).  Columns
    (left to right) for normal, exponential, and uniform distribution,
    rows for several relative widths $\tilde \sigma = \sigma / \langle
    E_W\rangle$ as indicated.  The spikes seen at an intermediate
    temperature, most notably for $S_1/S$, are a result of the switch of
    $E_\mathrm{cut}$ between $E_\mathrm{cut}^>$ and
    $E_\mathrm{cut}^<$.}\label{fig:cvgeta-comparison}
\end{figure}

We have introduced the mapping to an effective binary model, and we have
reviewed earlier (cf.\ \sref{sec:binary}) how this model is described
and solved using rate equations.  We will now compare its predictions
with the outcome of KMC simulations.  For each temperature and
distribution shape and width, we simulated 10 realizations as described
in \sref{sec:kmc}.  \Fref{fig:cvgeta-comparison} shows that overall,
agreement between the KMC results and the prediction of the effective
model is very good, for both the coverage and the efficiency.  Most
importantly, the temperature range of efficient reaction is reproduced
with very good accuracy in most circumstances---this is the truly
valuable information, compared with minor deviations in the efficiency
itself.
There is some discrepancy between KMC and effective model results in the
high-temperature tail for the normal and the exponential distribution.
The rate equation solution of the effective model describes hops between
\emph{any} two sites of the lattice, such that spatial correlations are
switched off entirely.  If we include such ``long hops'' in the KMC
simulations, the efficiency also increases (as checked in several test
runs, and as previously found and explained for the binary
system~\cite{wolff10-react-two-dimen-bin-diso}), and it then agrees even
better with the effective model results.

The system now shows a broad temperature window of high efficiency,
different from the homogeneous system, but also from the binary case, as
far as the slow decay to higher temperatures is concerned (cf.\
\fref{fig:cvgetabin}).  Likewise, we observe a smooth monotonic
transition from full coverage to an empty system, in stark contrast to
the binary case---this illustrates once more that the effective binary
model we map to changes its structure with temperature.
It is also evident that sample-to-sample fluctuations are a subordinate
effect throughout, even on the critical flanks of the efficiency and for
the long-tailed exponential distribution.  We will return to this issue
in \sref{sec:realizations}.

Of special interest is the range of validity of the mapping idea.  As
emphasized before, it relies on the presence of both types of sites,
shallow and deep, in the effective model.  This is no longer satisfied
at very low temperatures, when all sites are effectively deep ($S_1=0$),
and at very high temperatures, when all sites are effectively shallow
($S_2=0$).  Obviously, these limits are reached at less extreme
temperatures for narrower distributions (upper rows), and in the absence
of distribution tails, as exemplified by the uniform and (to lower
energies) the exponential distribution.
The effective model degenerates to a homogeneous system then, with the
binding energy given by the mean of the distribution.  For the figure,
we correspondingly replaced the numerical solution of the effective
binary model by the analytical results for the homogeneous case.

For very low temperatures, the examples studied here show essentially
full coverage and zero efficiency, which is trivially reproduced by the
homogeneous rate equation results.  The support of the exponential and
the uniform distribution is bounded to low energies.  Therefore, the
transition to the $S_1=0$ regime is not smooth, which manifests itself
in the discontinuous derivative of coverage and efficiency.
For high temperatures, the situation is more subtle.  The normal and the
exponential distribution, which both have tails to high energies, are
still accurately described at very high temperatures: For most of the
panels shown, $S_2=0$ is reached eventually, but only after the KMC
efficiency has vanished completely.  Up to this temperature, there are
still \emph{some} deep sites in the effective model owed to the
distribution tail (not visible in the plots due to limited resolution),
and they are sufficient to reproduce KMC results.  At still higher
temperatures, the effective homogeneous system trivially reproduces zero
coverage and efficiency.
For the uniform distribution, however, KMC results show a fast (but by
no means abrupt) decay of the efficiency with increasing temperature.
Due to the lack of high-energy tails (which could still provide a few
deep sites), the effective model now degenerates ($S_2=0$) at a
temperature low enough that KMC results still exhibit some efficiency.
For a very narrow distribution ($\tilde\sigma=10\%$) this happens so
early ($T\approx 18\ \mathrm K$) that the resulting effective model
(homogeneous with binding energy $\langle E_W\rangle$) still shows the
high-temperature flank seen in \fref{fig:cvgetahom}.  For all wider
distributions the switch to the degenerate effective model occurs at
temperatures where the homogeneous system has no efficiency left, while
the KMC results still have residual efficiency, most likely due to the
mere fact that there \emph{is} a distribution of different binding
energies (in part exceeding $\langle E_W\rangle$) and possibly some
spatial correlations.

\subsection{Tail shape and analytical expressions}
\label{sec:tails}

For a homogeneous system, the tail shape of the efficiency $\eta(T)$
(fairly symmetric to low and high temperatures) is well understood (cf.\
\sref{sec:homogeneous}): The efficiency decays exponentially with the
temperature in both cases, since all rates are thermally activated.
From the rate equation efficiency~\eref{eta-hom} one finds
\begin{equation}\label{hom-tails}
  \hspace{-\mathindent}
  \eta \simeq 2\frac{a}{f} = 2\frac{\nu}{f} \e^{-E_a/T}
  \quad\text{(low-$T$ tail)},
  \qquad
  \eta \simeq 2\frac{fa}{W^2} = 2\frac{f}{\nu} \e^{(2E_W-E_a)/T}
  \quad\text{(high-$T$ tail)},
\end{equation}
which mirrors the temperature bounds~\eref{Tlow} and~\eref{Tup}.  The
tail \emph{shapes} for the binary system are the same as for the
homogeneous system, since for each tail only reaction on one type of
site is important.

For continuously distributed binding energies, however, the situation is
different.  There are many similar binding energies acting almost but
not exactly the same (at a certain temperature).  This is reflected by
the slower decay of the efficiency.  We now use the mapping to the
effective binary model to derive an analytical expression for the
(low-temperature) tail shape.

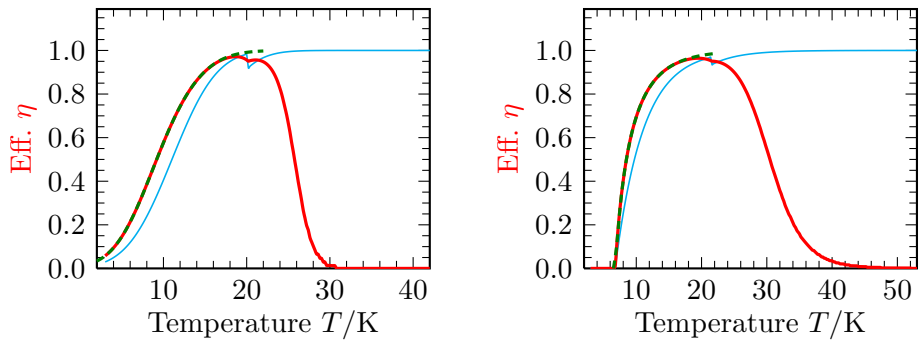
\begin{figure}
  \centering
  \tikzsetnextfilename{fig8a}%
  \begin{tikzpicture}
    \begin{axis}[
      ,width=\stepplotwidth,height=\stepplotheight
      ,xlabel=Temperature $T/\mathrm K$
      ,ylabel={\color{red}Eff.\ $\eta$}
      ,xmin=2,xmax=42,ymin=0,ymax=1.19
      ,xprec=0
      ,yprec=1
      ,minor x tick num=4,minor y tick num=3
      ]
      \pgfplotstableread{etap-loT-nrm-030.dat}\etap
      \pgfplotstableread{effbin-maxeta-nrm-30.dat}\EFF
      \addplot[supp graph,cyan] plot table[x index=0,y index=3] from \EFF;
      \addplot[main graph,red] plot table[x index=0,y index=2] from \EFF;
      \addplot[main graph,darkgreen,densely dashed]
      plot table[x index=0,y index=1] from \etap;
    \end{axis}
  \end{tikzpicture}\qquad%
  \tikzsetnextfilename{fig8b}%
  \begin{tikzpicture}
    \begin{axis}[
      ,width=\stepplotwidth,height=\stepplotheight
      ,xlabel=Temperature $T/\mathrm K$
      ,ylabel={\color{red}Eff.\ $\eta$}
      ,xmin=2,xmax=53,ymin=0,ymax=1.19
      ,xprec=0
      ,yprec=1
      ,minor x tick num=4,minor y tick num=3
      ]
      \pgfplotstableread{etap-loT-exp-030.dat}\etap
      \pgfplotstableread{effbin-maxeta-exp-30.dat}\EFF
      \addplot[supp graph,cyan] plot table[x index=0,y index=3] from \EFF;
      \addplot[main graph,red] plot table[x index=0,y index=2] from \EFF;
      \addplot[main graph,darkgreen,densely dashed]
      plot table[x index=0,y index=1] from \etap;
    \end{axis}
  \end{tikzpicture}
  \caption{Efficiency $\eta$ versus temperature $T$ in the effective
    binary model, evaluated using the numerical solution of rate
    equations (red, as in \fref{fig:cvgeta-comparison}) and as given by
    the analytical expression~\eref{eta-lowT} (green, dashed), for a
    normal distribution with $\tilde \sigma = 30\%$ (left), and for an
    exponential distribution with $\tilde\sigma = 30\%$ (right).  For
    reference, the fraction $S_1/S$ of shallow sites in the effective
    model is shown again (thin cyan line).}\label{fig:eta-lowT}
\end{figure}

As alluded to in \sref{sec:homogeneous}, the binary system exhibits a
plateau of the efficiency $\eta(T)$ between the two peaks of the
corresponding homogeneous systems in certain conditions.  More
precisely, one needs enough deep binding sites---depending on the
temperature, flux and the difference in the binding energies of the two
types of sites.
Following~\cite{wolff10-react-two-dimen-bin-diso}, we let
$T^\mathrm{eq}$ denote the temperature below which the random walk
length (on shallow sites), $\ell_{\mathrm{rw}} = \sqrt{a_1/W_1}$,
exceeds the average hopping length before encountering a
trap~\cite{montroll69, evans85}, $\ell_{\mathrm{trap}} \simeq
\sqrt{S/(\pi S_2)\cdot \ln S}$: At lower temperatures, particles
typically end in deep sites.
If $T^\mathrm{eq} > T_2^\mathrm{max}$ we find a plateau, with an
efficiency of~\cite{wolff10-react-two-dimen-bin-diso}
\begin{equation}\label{eta-p}
  \eta_{\mathrm p} \approx \frac{2}{1+S/S_1}.
\end{equation}
Now in the effective binary model, the energies and numbers of both
types of sites are functions of $E_{\mathrm{cut}}$ and thus depend on
the temperature $T$.
For the low temperature tail of all shown distributions, we have
sufficiently many deep sites $S_2$ in the effective model, such that the
condition $T^\mathrm{eq} > T_2^\mathrm{max}$ is satisfied---the
effective model (for the given temperature) features a plateau.  One
also finds that $T_1^\mathrm{max} < T < T_2^\mathrm{max}$, such that we
evaluate the model \emph{on} this plateau, and the formula~\eref{eta-p}
applies.
The fraction of shallow sites $S_1/S$ in the effective model is given by
the cumulative distribution function $\Phi(E_{\mathrm{cut}}) :=
\int_{-\infty}^{E_{\mathrm{cut}}} \rho(E_W) \,\dd E_W$.
Lastly, since we are in the low-temperature tail, we have
(cf.~\sref{sec:mapping-derivation}) $E_{\mathrm{cut}} =
E_{\mathrm{cut}}^> = T\ln(\nu/f) + \Delta E$, leading to
\begin{equation}\label{eta-lowT}
  \eta \approx \frac{2}{1+\Phi(T\ln(\nu/f)+\Delta E)^{-1}}
  \quad \text{(low $T$)}.
\end{equation}
This expression shows a much weaker dependence on temperature compared
with the homogeneous and (genuinely) binary cases with their exponential
decay.
It also explains that the broader tails of the efficiency do not
necessarily originate from tails of the underlying distribution
$\rho(E_W)$ (provided there still \emph{are} both deep and shallow
sites).  Rather, the decisive factor is that the mapping introduces a
$T$-dependent split into shallow and deep sites via the cutting energy
$E_\mathrm{cut}$---without thermally activated rates playing any role.
Moreover, this implies a lower temperature bound of efficient reaction
(where $\eta = 1/2$) given by
\begin{equation}
  \Phi(T^\mathrm{low} \ln(\nu/f)+\Delta E) = \frac{1}{3}.
\end{equation}
\Fref{fig:eta-lowT} shows that indeed the low-temperature
expression~\eref{eta-lowT} is extremely accurate up to intermediate
temperatures around the efficiency peak temperature.  This corresponds
to the fact that the plateau in the binary model breaks down only at
rather low fraction $S_2/S$ (depending on $E_{W_2}-E_{W_1}$).  We have
checked that these statements hold true for all parameters used in
\fref{fig:cvgeta-comparison}.

For the high-temperature tail, the situation is more subtle.  Here, the
effective binary model only has few deep sites, and they are far from
the mean energy $\langle E_W\rangle$ (cf.~\eref{E-W-eff}).  Even a fixed
such model has no efficiency plateau then, but a $T$-dependent
efficiency drop between the ``homogeneous peaks''~\cite[Figure
5]{wolff10-react-two-dimen-bin-diso}.  We do not have an illuminating
analytical expression for this dependence, wherefore the upper temperature
bound $T^\mathrm{up}$ remains inaccessible as well.

\subsection{Realization dependence}
\label{sec:realizations}

We first comment on the effect of the quenched nature of disorder at the
\emph{microscopic} level.  In any particular realization of the system,
sites with similar binding energy may live in a very different local
neighborhood.  Therefore, they can differ strongly in the occupancy $u$
and the reactivity $r$.  
This is very prominent in the raw simulation data, and the variability
was intentionally reduced in \fref{fig:steps-of-E} and
\fref{fig:steps-of-E2} by using a sliding average.
The site-to-site variability only vanishes in passing to the ensemble of
all realizations (or infinite system size $S\to\infty$), whereas purely
stochastic fluctuations decrease with increasing simulation time.
We have confirmed this distinction by comparison with KMC simulations in
which spatial correlations are suppressed (``long hops'' between all
sites are included, cf.\ \sref{sec:comparison}).  This indeed removes
the major part of the variability in occupancy and reactivity.

Interestingly, site-to-site variations are much more pronounced for the
reactivity than for the occupancy.  The reason for this is as follows:
Consider the system dynamics over a certain period of time, and we are
only concerned with effectively deep sites, where essentially all
coverage and reaction events are concentrated.
The number of such events on a given site is (to a good approximation)
Poisson-distributed, with a rate parameter depending on the local
surroundings.  Together with statistical fluctuations, this gives rise
to the variability seen in the reactivity $r$.
For the occupancy $u$, individual occupation times of a site are added
up and compared with the total time passed.  Since only a reaction event
empties the site (hopping and desorption from deep sites is negligible),
there are as many individual occupation times as there are reaction
events on this site.  This strongly anticorrelates the number of such
events with the length of individual occupation times---if particles
arrive more frequently, single occupation times are shorter.  Therefore
the number of reaction events (and hence, the reactivity) can strongly
differ between two sites of similar energy, yet the fraction of time
they are occupied (the occupancy) will differ far less.
Note that the reduced variability in the occupancy versus the reactivity
immediately translates to that of the total coverage versus the
efficiency between different realizations.

We now turn to this dependence of \emph{global} quantities on the
realization.  The overall system size in this article is large enough
not to expect a noticeable dependence of the coverage and the efficiency
on the realization.  This is confirmed in \fref{fig:cvgeta-comparison}
for the lower-temperature regime of both the coverage and the
efficiency.  For the high-temperature decay of the efficiency, however,
such a dependence is clearly seen in the vertical spread of symbols
referring to different realizations, both in the case of the normal and
the exponential distribution.  Somewhat counterintuitively, the
variability between realizations \emph{decreases} with increasing
disorder strength (width of the distribution).

The mapping to an effective model explains if and why we see significant
sample-to-sample variations of the efficiency.
As explained at the end of \sref{sec:tails}, for high temperatures the
effective binary model has few deep sites, far from the mean binding
energy.  We know (from both simulations and numerical solutions of the
rate equations), that in this regime the efficiency of the binary system
is very sensitive to the exact number of deep
sites~\cite{wolff10-react-two-dimen-bin-diso}.  This is perfectly
intuitive, since there are so few of them, yet they are very important
for the reaction.
Applying the mapping to \emph{different realizations} of the finite
continuous-distribution system, the number of \emph{effectively} deep
sites also varies, and because there are few in any case, the variations
relatively matter a lot.  The sensitivity of the effective binary model
to their number (at fixed $T$) thus explains the realization dependence
of the KMC efficiency of the continuous system
(\fref{fig:cvgeta-comparison}), and why it only shows on the
high-temperature flank.  Moreover, it is more pronounced for narrower
distributions of the binding energy, since steeper flanks of the PDF
lead to larger relative variations in the small number of effectively
deep sites.
The coverage is already very small in this regime, such that its
realization dependence is not visible in \fref{fig:cvgeta-comparison}.

It is an interesting feature that, though part of a nominally large
system, the smallness of one crucial component (the number of deep
sites) is enough to imply a strong realization dependence of a key
quantity such as the efficiency.  This constitutes an \emph{effective
  small-system regime}, in the sense that the realization dependence
will still vanish as usual upon increasing the total system size $S$.
In this context, the mapping to an effective model concisely explains
that depending on the temperature, we are in different regimes as to the
effect of disorder.
The asymmetry between shallow and deep sites (i.e., why is there no
strong sensitivity when there are only few of the former?) is easily
resolved.  At temperatures so low that there are very few effectively
shallow sites only, $S_1/S\ll1$, application of the plateau
formula~\eref{eta-p} (as justified in \sref{sec:tails}) yields
$\eta_\mathrm{p} \approx 2 S_1/S \ll 1$.  Therefore, whatever
sample-to-sample variability there is in the efficiency cannot be seen
in \fref{fig:cvgeta-comparison}.  On the other hand, the sensitivity
of the coverage to the realization is much weaker anyway, as shown
above.

\section{Conclusions}
\label{sec:conclusions}

We have studied the steady state of a two-dimensional diffusion-limited
reaction model, with disordered binding energies drawn from a continuous
distribution.  Sites in this model play one of two distinct roles, as we
have verified in simulations: If the binding energy is low enough, sites
provide mobility of particles to traverse the surface.  If their binding
energy is strong enough, they instead provide coverage by trapping
particles for a long time.  As a result, we can map the
continuous-distribution model to an \emph{effective binary model} of
these shallow and deep sites, which is well understood and easily
solved.  The precise form of the mapping has been derived heuristically
and does not depend on any fitting parameters.  The model yields results
for the coverage and the reaction efficiency which are in good agreement
with simulations.
Compared with the case of discrete distributions studied before, the
model shows a markedly different behavior, with the temperature range of
efficient reaction broadened and the tails decaying much slower.
The mapping explains this slower decay for low temperatures, as well as
the sample-to-sample fluctuations found for the high-temperature decay
of the reaction efficiency.

As discussed in the introduction, the particular model studied here is
paradigmatic for applications in astrophysics and in heterogeneous
surface catalysis.  Moreover, the existence of a simple mapping from a
highly complex to a simple effective model is of great theoretical
value.  The explanations we provide in terms of microscopic processes
can hopefully serve as a recipe to find similar relations for other
types of disordered systems.

\ack We thank Thomas Nattermann for useful suggestions.  This work was
supported by Deutsche Forschungsgemeinschaft within SFB/TR-12
\emph{Symmetries and Universality in Mesoscopic Systems} and the
Bonn-Cologne Graduate School of Physics and Astronomy, and by the
US-Israel Binational Science Foundation.  IL gratefully acknowledges
financial support by the Minerva foundation.

\section*{References}
\bibliography{jrnlabrv,myrefs}

\end{document}